\documentclass[a4paper,12pt]{article}
\pdfoutput=1 
\usepackage[utf8]{inputenc}
\usepackage{float}
\usepackage{array}
\usepackage{appendix}
\usepackage{lipsum}
\usepackage{graphicx}
\usepackage{amsmath,amsthm,amsfonts,amssymb}
\usepackage{graphicx}
\usepackage[english]{babel}
\usepackage{color}
\usepackage{subcaption}   
\usepackage{tensor}
\usepackage{esint}
\usepackage[dvips]{epsfig}
\usepackage[dvips]{graphicx}
\usepackage{float}
\usepackage{units}
\usepackage{textcomp}
\usepackage{mathrsfs}
\usepackage{amsmath}
\usepackage[makeroom]{cancel}
\usepackage{amssymb}
\usepackage{amsbsy}
\usepackage{amsfonts}
\usepackage{amssymb,mathrsfs,xcolor}
\usepackage{esint}
\usepackage{array}
\usepackage{graphicx}
\usepackage{jcappub} 
\usepackage{bigints}
\usepackage[T1]{fontenc} 

\newcommand{\be}{\begin{equation}}
\newcommand{\ee}{\end{equation}}
\newcommand{\bea}{\begin{eqnarray}} 
\newcommand{\eea}{\end{eqnarray}}
\newcommand{\bes}{\begin{subequations}}
\newcommand{\ees}{\end{subequations}}

\newcommand{\cD}{{\cal D}}
\newcommand{\sech}{{\rm sech}}

\title{\boldmath New brane-like solutions in modified four-dimensional Einstein-Gauss-Bonnet gravity}

\author[a]{D. Bazeia,}
\author[b]{R. Menezes,}
\author[a]{A. Yu. Petrov,}
\author[a,1]{P. J. Porfírio \note{Corresponding author.}}

\affiliation[a]{Departamento de Física, Universidade Federal da Paraíba, Caixa Postal 5008, 58051-970, João Pessoa, Paraíba,  Brazil.}

\affiliation[b]{Departamento de Ci\^{e}ncias Exatas, Universidade Federal da Para\'{\i}ba, \\ 58927-000, Rio Tinto, PB, Brazil.}

\emailAdd{bazeia@fisica.ufpb.br}
\emailAdd{rmenezes@dcx.ufpb.br}
\emailAdd{petrov@fisica.ufpb.br}
\emailAdd{pporfirio@fisica.ufpb.br}

\abstract{We investigate solutions of a new $4D$ Einstein-Gauss-Bonnet gravity ($4D$ $EGB$). We first describe the bulk vacuum solution, then we add a massive probe scalar field, and we follow considering a self-interacting scalar field which acts as a source to support thick brane solutions in the four-dimensional $EGB$ scenario with a single extra dimension of infinite extent. We illustrate our results with some distinct brane-like configurations engendering controllable thickness. 
It is noteworthy that such configurations are simultaneous solutions in both versions of the modified theory of gravity, the original Glavan and Lin formulation and the regularized $4D$ $EGB$.}

\begin{document}
\maketitle
\flushbottom


\section{Introduction}
\label{sec:intro}

Fundamental discoveries of last years, such as late-time cosmic acceleration \cite{Cos1,Cos2}, detection of gravitational waves (see f.e. \cite{LIGO}), and obtaining the image of the shadow of a black hole (see f.e. \cite{Akiyama}), called strongest attention to studies of gravity. The known difficulties faced by general relativity (GR) (namely, its non-renormalizability, at the quantum level, and the impossibility of explaining the accelerated expansion of the Universe and formation of large-scale structures in the Universe, without postulating the existence of an amount of unseen matter and energy)  clearly elevate the importance of search for its consistent extension. Several modified gravity models have been considered in this context, those ones involving modification of geometric sector only (f.e. $f(R)$ gravity, see \cite{Sotiriou} for a review) or those ones including extra fields treated as ingredients of a complete gravity description rather than as matter fields (f.e. scalar-tensor or vector-tensor gravities). An excellent review on applying modified gravity models within the cosmological context is presented in \cite{OdNoj, Nojiri:2017ncd, Odintsov:2023weg}, and a wide list of classes of extended gravity models is presented in \cite{ourrev}.

Among various modifications of gravity, one of the very interesting examples is given by the four-dimensional Einstein-Gauss-Bonnet $(EGB)$ gravity \cite{newGB}. This theory possesses the following advantage -- it allows to involve only equations of motion of second order, thus avoiding the presence of ghosts, and introducing, at the same time, a richer structure in the action. Actually, this theory involves the additive Gauss-Bonnet (GB) term with a special rescaling of its coupling constant, $\alpha\rightarrow{\alpha}/{({\cal D}-4)}$, and, simultaneously, taking the limit $\mathcal{D}\rightarrow 4$, with ${\cal D}$ being the space-time dimension; this procedure converts the topological term to a physical one and effectively introduce some kind of dimensional reduction similar to what happens in quantum field theory (cf. \cite{Siegel}). In fact, the origin of the resultant non-trivial GB term in the field equations lies, in part, motivated by quantum corrections to the energy-momentum tensor \cite{Duff:1993wm, Fernandes:2022zrq}. 
The procedure considered by Glavan and Lin \cite{newGB} was proposed by Tomozawa in \cite{Tomozawa}, in which he considered one-loop radiative corrections to GR. In \cite{Cognola}, the authors reformulated the Tomozawa proposal just uplifting the classical action to $D$ dimensions and carrying out the entropic dimensional reduction to $D=4$. Getting back to four-dimensional $EGB$ theory, it should be mentioned some formal inconsistencies in taking the limit $\mathcal{D}\rightarrow 4$ at the level of field equations, since the regularization process is in contradiction to the Lovelock theorem (see \cite{Gurses1, Gurses2, Delhom1, Delhom2, Maha} for more details). Nevertheless, this formal issue was circumvented by carrying out the dimensional reduction procedure at the level of action (see \cite{Aoki1, Aoki2, Hennigar}), then rendering it a non-pathological theory. Actually, following this approach, it can be seen in a more transparent way that $EGB$, upon Kaluza-Klein dimensional reduction down to four dimensions, reduces to a scalar-tensor theory of gravity \cite{Lu, Fernandes2, Hint, Easson:2020mpq, Kobayashi}; thereby possessing a scalar field degree of freedom in addition to the gravitational ones. A variety of exact solutions has been obtained within the framework of four-dimensional $EGB$ \cite{newGB, Zhang, Fernandes3, Wei, Kumar, Kumar2, Qiao, Yang:2020jno, Lin:2020kqe, Bravo, Konoplya, Konoplya2, Godani, Papnoi, Jaryal, Malafarina:2020pvl, Doneva:2020ped, Charmousis:2021npl,Gammon:2023uss}.

One of the natural tasks here is to find other solutions, and this will be the main concern of the present work. In particular, we will also focus on the construction of thick brane solutions in this new environment. Our motivation here is that the presence of thin and thick brane-like configurations \cite{Randall1,Maartens} in the four-dimensional $AdS$ bulk with the inclusion of the GB contribution may also bring novelties concerning old results on black holes \cite{2brane1,2brane2,Carol}, an issue that may lead us to ask new questions about strong gravity on thick branes. Numerous studies addressing brane-like solutions have been carried out in the context of other modified theories of gravity. For example, thin and thick branes solutions in higher-dimensional $f(R)$ gravity have been found in \cite{FRB1,fr1,fr2,FRB2,fr3} and in references therein; see also \cite{Kallosh, Brito, Ahn, Karn} for issues within the context of supergravity.  Another intrinsic motivation is rooted in the realm of quantum gravity, specifically within the framework of string theory, where brane-like solutions naturally emerge, highlighting the crucial role played by $p$-branes in this context (see \cite{Stelle, Skenderis1999} for a detailed review). In particular, it has been shown in \cite{Gibbons:1993sv} that a broader class of $p$-branes interpolates between $AdS_{p+2}\times S^{\mathcal{D}-(p+2)}$, where $S^{n}$ is a $n$-sphere embedded in a $\mathcal{D}$-dimensional spacetime, in the near-horizon geometry limit
and a $\mathcal{D}$-dimensional Minkowski space in the asymptotic limit (at spatial infinity). 
A holographic description was inspired by geometrical features of such branes in the near-horizon limit, providing a powerful theoretical framework for the study of strongly coupled gauge theories (see f.e. the original references \cite{Maldacena, EWitten, KlebWitten}). On the other hand, given the mathematical difficulties associated with higher-dimensional gravity theories \cite{Carlip}, lower-dimensional theories are frequently employed to investigate fundamental issues in quantum gravity \cite{Carlip} or to mimic certain four-dimensional setups in laboratory settings \cite{Novello, Visser}. In braneworld scenarios, akin to the Dvali-Gabadadze-Porrati (DGP) approach \cite{Dvali}, $(2+1)$-dimensional induced gravity on a $2$-brane embedded in a flat four-dimensional bulk has been explored in the literature \cite{CBrito1, CBrito2, Porfirio1}. Their holographic properties have also been explored within the context of AdS$_3$/CFT$_2$ duality \cite{Strominger1,Strominger2}; as well as in lower dimensions through the AdS$_2$/CFT$_1$ correspondence \cite{Cvetic1}.

In view of the above considerations, we organize the work as follows: in  
Section 2, we write down the action of our theory and the modified Einstein equations. After, in Section 3 we describe the main results. There we first deal with the vacuum solution and then find the results for the massive probe scalar field in this gravitational background. We also investigate and solve the modified Einstein equation for the scalar field acting as a source field to generate thick brane solutions in the bulk $AdS_4$, in a way similar to the case of thick brane in the $AdS_5$ geometry with a single extra spatial dimension of infinite extent. We end the investigation in Section 4, where we summarize our results and add some new possibilities of continuation of the work.


\section{Four-dimensional Einstein-Gauss-Bonnet gravity}
\label{general}

Recently, a new four-dimensional $EGB$ gravity was proposed \cite{newGB}. Its action is initially defined in a $\mathcal{D}$-dimensional space, and then one takes the limit $\mathcal{D}\to 4$. Thus, $EGB$ action looks like 
\bea
S=\int d^{\mathcal{D}}x\sqrt{-g}\left(\frac{1}{2\kappa^2}R-\frac{\alpha}{{\cal D}-4}{\cal G}+{\cal L}_m\right),
\label{GBaction}
\eea
where $\kappa^2=8\pi G$ is related to Newton's constant, $\mathcal{G}=\left(R^2 - 4 R_{\mu\nu}R^{\mu\nu}+R_{\mu\nu\alpha\beta}R^{\mu\nu\alpha\beta}\right)$ is the GB scalar invariant, and the other geometrical entities are: the Ricci scalar $\left(R=g^{\mu\nu}R_{\mu\nu}\right)$, the Ricci tensor $\left(R_{\mu\nu}\right)$ and the Riemann tensor $\left(R^{\mu}_{\,\,\,\nu\alpha\beta}\right)$. The matter sources are minimally coupled with the metric through the Lagrangian, $\mathcal{L}_{m}$.  Conversely, to the usual $EGB$ gravity, this novel $4D$ version of the theory is achieved by rescaling the GB coupling constant to $\dfrac{\alpha}{\mathcal{D}-4}$, where $\alpha$ is a dimensionless constant. It has been shown in \cite{newGB} that, by taking the limit $\mathcal{D}\rightarrow 4$, the theory provides non-trivial contributions to the gravitational field equations stemming from the GB action, as we shall see in what follows.

By varying the action (\ref{GBaction}) with respect to the metric, we obtain the following equations of motion
\bea
\frac{1}{\kappa^2}G^{\mu}_{\nu}+\frac{2\alpha}{{\cal D}-4}\mathcal{H}^{\mu}_{\nu}=T^{\mu}_{\nu},
\label{fe}
\eea
where $G^{\mu}_{\nu}$ is the usual Einstein tensor, 
\begin{equation}
\mathcal{H}^{\mu}_{\nu}=\left(-2R^{\mu\alpha}_{\phantom{\mu\alpha}\rho\sigma}R^{\rho\sigma}_{\phantom{\rho\sigma}\nu\alpha}+4R^{\mu\alpha}_{\phantom{\mu\alpha}\nu\beta}R^{\beta}_{\alpha}+4R^{\mu}_{\alpha}R^{\alpha}_{\nu}-2RR^{\mu}_{\nu}+\frac{1}{2}{\cal G}\delta^{\mu}_{\nu}\right)
\end{equation}
is the contribution stemming from the Gauss-Bonnet term and 
\begin{equation}
    T_{\mu\nu}=-\frac{2}{\sqrt{-g}}\frac{\delta(\sqrt{-g}\mathcal{L}_{m})}{\delta g^{\mu\nu}},
\end{equation}
is the stress-energy tensor of the matter sources.

Taking the trace of $\mathcal{H}^{\mu}_{\nu}$, one finds the general relation
\begin{equation}
    \mathcal{H}^{\mu}_{\mu}\equiv \mathcal{H}=\frac{\mathcal{D}-4}{2}\mathcal{G},
\end{equation}
which, due to the constant factor $(\mathcal{D}-4)$ accompanying $\mathcal{G}$ in the above equation, provides a finite contribution to the trace equation of (\ref{fe}). However, the question as to whether this property holds or not for the tensorial equation (\ref{fe}) is a key point. It has been addressed in several works,  see f.e. \cite{Fernandes:2022zrq,Gurses1} and the original one itself \cite{newGB}. To shed more light on this, we call attention to the fact that  $\mathcal{H}_{\mu\nu}$ can be split into two pieces \cite{Gurses1}, namely,
\begin{equation}   \mathcal{H}_{\mu\nu}=2\left(\mathcal{L}_{\mu\nu}+\mathcal{Z}_{\mu\nu}\right),
\end{equation}
where
\begin{eqnarray}
\mathcal{L}_{\mu\nu}=C_{\mu\alpha\beta\gamma}C_{\nu}^{\phantom{a}\alpha\beta\gamma}-\frac{1}{4}g_{\mu\nu}C_{\alpha\beta\gamma\sigma}C^{\alpha\beta\gamma\sigma},  
\end{eqnarray}
with $C_{\mu\alpha\beta\gamma}$ being the Weyl tensor and
\begin{eqnarray}   \nonumber\mathcal{Z}_{\mu\nu}=\frac{\left(\mathcal{D}-4\right)\left(\mathcal{D}-3\right)}{\left(\mathcal{D}-1\right)\left(\mathcal{D}-2\right)}\bigg[&-&\frac{2\left(\mathcal{D}-1\right)}{\left(\mathcal{D}-1\right)}C_{\mu\rho\nu\sigma}R^{\rho\sigma}-\frac{2\left(\mathcal{D}-1\right)}{\left(\mathcal{D}-1\right)}R_{\mu\lambda}R^{\lambda}_{\nu}+\frac{\mathcal{D}}{\left(\mathcal{D}-2\right)}R_{\mu\nu}R\\
    &+&\frac{1}{\left(\mathcal{D}-2\right)}g_{\mu\nu}\left((\mathcal{D}-1)R_{\alpha\beta}R^{\alpha\beta}-\frac{\left(\mathcal{D}+2\right)}{4}R^2\right)\bigg].
\end{eqnarray}
It is straightforward to see that $\mathcal{Z}_{\mu\nu}$ is well-defined when taking the limit $\mathcal{D}\to 4$, while $\mathcal{L}_{\mu\nu}$ is not, as discussed in \cite{Fernandes:2022zrq}. In fact, for $\mathcal{D}\leq 4$, $\mathcal{L}_{\mu\nu}$ vanishes completely, therefore it cannot be written in the form   $\mathcal{L}_{\mu\nu}=\left(\mathcal{D}-4\right)\mathcal{Y}_{\mu\nu}$, for a generic metric. Yet, for particular classes of metrics, such as $\mathcal{D}$-dimensional Friedmann-Robertson-Walker (FRW), $\mathcal{D}$-dimensional maximally symmetric and spherically symmetric spacetimes, the limit $\mathcal{D}\to 4$ is well-behaved. Some other subtleties have also been discussed in the literature,  see e.g. \cite{Gurses1}, e.g., a well-defined perturbative scheme for the field equations, which points out that the Glavan and Lin procedure is not well-defined \cite{newGB}. Fortunately, there are covariant alternative regularizations at the action level to circumvent this problem, namely: conformal regularization \cite{ Fernandes2} and regularization via Kaluza-Klein reduction \cite{Lu, Hint, Kobayashi}. After performing the regularization, the remaining action is such that it belongs to a particular Horndeski class of scalar-tensor theories \cite{Horndeski, Kob}. In the following, we obtain the field equations using the Glavan and Lin method \cite{newGB}; however, we also present a completely well-defined version of $4D$ $EGB$ 
(we shall call it regularized $4D$ $EGB$), which is based on a Kaluza-Klein regularization at the action level in Appendix \ref{app}. 
Remarkably, some specific metrics, like those aforementioned (FRW, $\mathcal{D}$-dimensional maximally symmetric and spherically symmetric spacetimes \cite{Fernandes:2022zrq}) are solutions for both versions of the theory: the original formulation and the regularized $4D$ $EGB$.

As a first attempt, we can consider the toy model of the brane-like metric where one of the axes (say $w$) plays the role of the extra dimension. Then, the metric ansatz reads
\bea
ds^2=e^{2A(w)}\eta_{ab}dx^adx^b-dw^2,
\label{metric}
\eea
where $\eta_{ab}$ is the $(\mathcal{D}-1)$-dimensional Minkowski metric 
and $A(w)$ is the warp function.
Recalling that we pick the following convention: small Greek letters to label bulk coordinates $(\mu,\nu=0,1...D)$, with $D=\mathcal{D}-1$, and small Latin letters to label ``parallel'' coordinates to the brane $(a,b=0...D-1)$. For the metric (\ref{metric}), the relevant non-trivial geometrical quantities are
\bea
\Gamma^a_{bD}&=&A^{\prime}\delta^a_b;\quad\, \Gamma^D_{bc}=\eta_{bc}e^{2A}A^{\prime};\nonumber\\
R^a_{\phantom{a}bcd}&=&(\delta^a_c\eta_{bd}-\delta^a_d\eta_{bc})(A^{\prime})^2e^{2A};\quad\, R^{D}_{\phantom{D}bDd}=\eta_{bd}e^{2A}(A^{\prime\prime}+(A^{\prime})^2);\nonumber\\
R_{bd}&=&\eta_{bd}e^{2A}[({\cal D}-1)(A^{\prime})^2+A^{\prime\prime}];\quad\, R_{DD}=-({\cal D}-1)[A^{\prime\prime}+(A^{\prime})^2];\nonumber\\
R&=&({\cal D}-1)[2A^{\prime\prime}+{\cal D}(A^{\prime})^2].
\eea
Here, the prime stands for derivative with respect to $w$.
These expressions will be used to construct our equations. Notice that the components of the Einstein tensor can be decomposed into two pieces, namely,
\bea
G^a_b=(2-{\cal D})\delta^a_b[(\frac{{\cal D}-1}{2})(A^{\prime})^2+A^{\prime\prime}];\quad\, G^D_D=(\cD-1)(\frac{2-\cD}{2})(A^{\prime})^2.
\label{GG}
\eea
Further, 
\bea
R_{a\nu}R^{\nu c}&=&\delta_a^c[(\cD-1)(A^{\prime})^2+A^{\prime\prime}]^2, \quad\, R_{D\nu}R^{\nu D}=(\cD-1)^2(A^{\prime\prime}+(A^{\prime})^2)^2;\nonumber\\
R^{a\alpha}_{\phantom{a\alpha}b\beta}R^{\beta}_{\alpha}&=&\delta^a_b[D^2(A^{\prime})^4+(3D-1)A^{\prime\prime}(A^{\prime})^2+
D(A^{\prime\prime})^2],\nonumber\\
R^{D\alpha}_{\phantom{D\alpha}D\beta}R^{\beta}_{\alpha}&=&D^2(A^{\prime})^4+D(D+1)A^{\prime\prime}(A^{\prime})^2+D(A^{\prime\prime})^2;
\nonumber\\
R^{a\mu\nu\lambda}R_{b\mu\nu\lambda}&=&\delta^a_b[2D(A^{\prime})^4+2(A^{\prime\prime})^2+4 A^{\prime\prime}(A^{\prime})^2];\nonumber\\
R^{D\mu\nu\lambda}R_{D\mu\nu\lambda}&=&2D[(A^{\prime})^4+2A^{\prime\prime}(A^{\prime})^2+(A^{\prime\prime})^2];\nonumber\\
\mathcal{G}&=&D(D-1)(D-2)[(D+1)(A^{\prime})^4+4A^{\prime\prime}(A^{\prime})^2].
\label{G}
\eea

Substituting Eqs.(\ref{GG}) and (\ref{G}) into Eq.(\ref{fe}) and, after that, taking the limit $\mathcal{D}\rightarrow 4$, we find 
\begin{equation}
    \begin{split}
\kappa^2 T^{a}_{b}&=\left[-3(A^{\prime})^2 -2 A^{\prime\prime}+2\alpha\kappa^2 \left(3(A^{\prime})^4 + 4A^{\prime\prime}(A^{\prime})^2 \right)\right]\delta^{a}_{b};\\
\kappa^2 T^{3}_{3}&=-3(A^{\prime})^2+6\alpha\kappa^2 (A^{\prime})^4, 
    \end{split}
    \label{kj}
\end{equation}
where now $a,b=0...2$ and $D=3$.  The same field equations are obtained via regularized Kaluza-Klein reduction, as we explicitly show in  Appendix \ref{app}. 
As a consequence, the solutions for the ansatz \eqref{metric} to the field equations for the regularized $4D$ $EGB$ match those found in the Glavan and Lin formulation.


\section{Results}
\label{FEI}

Let us now consider some specific situations, where we can investigate and find results of current interest. We shall first deal with the bulk vacuum solution, then include a massive probe scalar field, and then make the scalar field to support self-interaction, taking it as a source to study its ability to support brane-like configurations with a single extra dimension of infinite extent. 

\subsection{The bulk vacuum solution}

We now consider the absence of source fields. Different from Einstein's theory 
without cosmological constant, $4D$ $EGB$ gravity supports a non-trivial solution in the presence of a Randall-Sundrum-like thin brane in $(3+1)$-dimensional bulk \cite{Randall1, Maartens} 
with no need for a non-trivial cosmological constant. 
In our case, the solution describes two copies of $AdS_4$ glued together by a boundary, which one calls 2-brane. In this scenario,
we follow the standard Israel junction conditions \cite{Israel:1966rt} to proceed with the gluing process, which is reflected by a $\delta$-function source in the stress-energy tensor, i.e, $T^{a}_{b}= \delta(w)\delta^{a}_{b}$, where the brane is placed at $w=0$ (of course, $T^{3}_{3}=0$,
 and we are taking the brane tension to be equal to one). In this way,  by solving the field equations (\ref{kj}), one obtains the following solution 
\begin{equation} \label{A_w}
A(w)=-\dfrac{1}{\sqrt{2\alpha}\kappa}|w|,
\end{equation}
which leads to the bulk metric
\begin{equation}
    ds^2=e^{-\sqrt{\frac{2}{\alpha}}\frac{|w|}{\kappa}}\eta_{ab}dx^{a}dx^{b}-dw^2,
    \label{bulk}
\end{equation}
where $\alpha$ should be a positive constant. 
The induced metric of the brane is given by the boundary condition: $g_{\mu\nu}(x^a,w)|_{w=0}=\eta_{ab}\delta^{a}_{\mu}\delta^{b}_{\nu}$.

Note that the bulk metric \eqref{bulk} describes 
two copies of $AdS_{4}$ spacetime 
with a Minkowski brane localized at $w=0$. The particular $|w|$ term reflects the $\mathrm{Z}_{2}$-symmetry around the brane. In fact, in order to see that in more detail, let us perform the following coordinate transformation: $e^{-\sqrt{\frac{1}{2\alpha}}\frac{|w|}{\kappa}}=\dfrac{r_{_{AdS}}}{z}$, where $r_{AdS}<z<+\infty$ and $r_{_{AdS}}^{2}=-\dfrac{3}{\Lambda}$, with $\Lambda<0$ and $r_{AdS}$ being the $AdS_4$ radius. 
Therefore, this geometry describes an $AdS_4$ spacetime  away from the Minkowski brane. The conformal boundary of $AdS$ is by definition located at $z=0$, where the warped factor blows up. Then, the bulk metric can be written in the upper half plane representation of $AdS_{4}$ \cite{Witten} by using the new coordinate $z$, namely,
\begin{equation}
    ds^2=\frac{r^{2}_{AdS}}{z^2}\left(\eta_{ab}dx^{a}dx^{b}-dz^2\right),
    \label{vc}
\end{equation}
in which the effective cosmological constant is defined by $\Lambda=-\dfrac{3}{2\kappa^2 \alpha}<0$. 
In this new coordinate system, the brane lives on a slice placed at $z=r_{AdS}$ and the full spacetime consists of gluing together two copies of $AdS_{4}$  across the brane. Therefore, the vacuum solution (\ref{vc}) emerges as a result of the non-trivial contribution of the GB term for the field equations in four dimensions, so that the physical effect appears as an effective negative cosmological constant which, in turn, is sourced by the coupling constant $\alpha$.  For the sake of convenience, let us proceed with a further coordinate transformation: $u=\dfrac{r_{AdS}^{2}}{z}$. By doing so, the line element \eqref{bulk} takes the form 
\begin{equation}
    ds^2 = \frac{u^2}{r_{AdS}^2}\eta_{ab}dx^{a}dx^{b}-\frac{r_{AdS}^{2}}{u^2}du^2,
    \label{lk}
\end{equation}
where the brane is now placed at $u=r_{AdS}$.

\subsection{Massive probe scalar field in Gaussian null coordinates}

We here investigate a massive probe scalar field in the background (\ref{vc}). The scalar field action is given by
\begin{equation}
    S[\Phi]=-\frac{1}{2}\int d^{4}x\sqrt{-g}\bigg(g^{\mu\nu}\partial_{\mu}\Phi\partial_{\nu}\Phi-m^2 \Phi^2\bigg).
\end{equation}
As it is well known $AdS_{4}$ spaces are not compact manifolds, in other words, they possess infinite volume. Hence, {\it a priori}, the above action can diverge. 
In the $4D$ analog of the Randall-Sundrum metric \eqref{bulk} this divergence problem is circumvented since the original boundary is shifted  from $z=0$ to $z=r_{AdS}$, where the $2$-brane lives in. By varying the action with respect to $\Phi$, one gets
\begin{eqnarray}
\nonumber 0&=&\int d^{3}x\int_{r_{AdS}}^{\infty} dz\left\{\partial_{\mu}\left(\sqrt{-g}g^{\mu\nu}\delta\Phi\partial_{\nu}\Phi\right)-\left[\frac{1}{\sqrt{-g}}\partial_{\mu}\left(\sqrt{-g}g^{\mu\nu}\partial_{\nu}\Phi\right)+m^2 \Phi\right]\sqrt{-g}\delta\Phi\right\}\\
 &=&\!-\!\int \!d^{3}x\!\int_{r_{AdS}}^{\infty}\!\!\! dz \!\left[\frac{1}{\sqrt{-g}}\partial_{\mu}\!\left(\sqrt{-g}g^{\mu\nu}\partial_{\nu}\Phi\right)\!+\!m^2 \Phi\right]\!\sqrt{-g}\delta\Phi+\nonumber\\ &+& \int d^{3}x \sqrt{-h}\,n^{\mu}\partial_{\mu}\Phi\delta\Phi \bigg{|}_{z=r_{AdS}}^{z=\infty},
\end{eqnarray}
where we have defined: $h$ is the determinant of the induced metric, $h_{\mu\nu}=g_{\mu\nu}+n_{\mu}n_{\nu}$, defined at the brane, and $n^{\mu}$ is the unit normal vector to the brane. Note that the first integral in the former equation is just the l.h.s. of the Klein-Gordon equation; consequently, it vanishes on-shell, and the second term is a boundary one, which must vanish under certain special boundary conditions in order to ensure a well-defined variation principle. Therefore, upon fixing the boundary conditions, one just needs to solve the Klein-Gordon equation in the bulk. 

 It should be noted that the metric (\ref{lk}) in Poincare coordinates is singular at the horizon ($u=0$), so it is more instructive to define a coordinate system where the metric is completely non-singular at the horizon. The suitable set of coordinates for this task is the so-called Gaussian null coordinates \cite{Friedrich, Kunduri}, where such new coordinates in the neighborhood of the horizon are, explicitly,
$(v,r,y^{A})$, with $v$ is the incoming null coordinate, $y^{A}$, with $A=1,2$, parameterizes the $2$-dimensional spatial section of the whole space, $r$ is the radial coordinate and, in particular, $0\le y^{(1)}=\eta<+\infty$. The Gaussian null coordinates are related to the Poincare ones by the following relation:
\begin{equation}
    u=r\cosh{\eta},\,\,\,\, t=\left(v+\frac{1}{r}\right)r_{AdS}^2 ,\,\,\,\, x^{A}=\left(\frac{\mu^{A}\tanh{\eta}}{r}\right)r_{AdS}^{2},
\end{equation}
where we have defined $\mu^{A}$ as a quantity to parameterize the $2$-dimensional section of the full space and must fulfill the requirement $\mu^{A}\mu^{A}=1$.

The metric in Gaussian null coordinates looks like 
\begin{equation}
ds^{2}_{N}=r_{AdS}^2\left[\cosh{\eta}\left(r^{2}dv^2 -2dv dr\right)-d\eta^{2} -\sinh^{2}{\eta}\,d\theta^2\right],
\label{null}
\end{equation}
where $\theta$ is the angular coordinate. The horizon is placed at $r=0$, then the metric is regular at the horizon.

\begin{figure}[ht]
    \centering
    \includegraphics[width=0.6\textwidth]{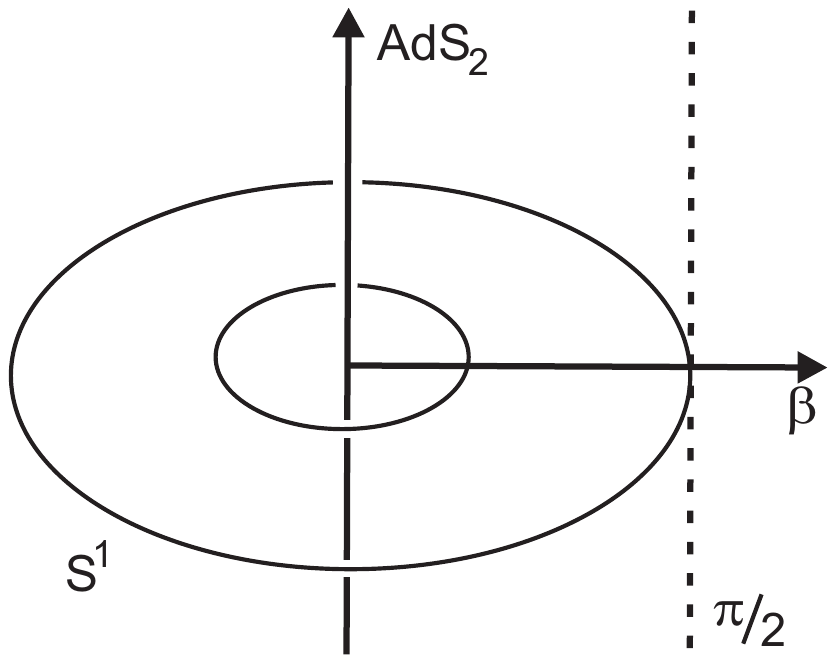}
    \caption{This plot schematically displays $AdS_{4}$ topology. $S^{1}$ represents circles of constant radii. The dashed vertical axis placed at $\beta=\pi/2$ stands for the boundary localization. At $\beta=0$, we have the vertical axis which represents $AdS_{2}$.
}
    \label{fig1}
\end{figure}

The bulk ($AdS_{4}$) metric, as expressed in terms of Gaussian null coordinates, Eq.(\ref{null}), can be viewed as a warped product of $AdS_{2}$
with a $2$-dimensional hyperbolic space, $\mathcal{H}^2$. In these coordinates, the 
$AdS_{4}$ boundary is located at $\eta=+\infty$. From the another perspective,
 using the conformal symmetry of $AdS$ spaces, by making the coordinate transformation $\tan{\beta}=\sinh{\eta}$,  the boundary can be brought to a finite point,
i.e, $\beta=\pi/2$. Explicitly, we have
\begin{equation}
ds^{2}_{N}=\frac{r_{AdS}^{2}}{\cos{\beta}^2}\left[r^{2}dv^2 -2dv dr-d\beta^2 -\sin^{2}{\beta}\,d\theta^2\right],
\label{eqw}
\end{equation}
 where $0\le \beta < \pi/2$. In this description, the $AdS_{4}$ space can be seen as a warped geometry of $AdS_{2}\times S^{2}$, whose topology 
of the conformal boundary (located at $\beta=\pi/2$) is simply $AdS_{2}\times S^{1}$, which is locally equivalent (isometric) to $\mathbf{R}^{3}$. Fig. \ref{fig1} displays the $AdS_{4}$ topology. Note that the vertical axis, located at $\beta=0$, describes the section of the full space corresponding to $AdS_{2}$, whilst the bulk of the ``cylinder'' represents $AdS_{4}$.

Having obtained a non-singular description of the $AdS_{4}$ metric, we now solve the Klein-Gordon equation on this background (\ref{eqw}), which is explicitly given by
\begin{equation}
    \left(\Box +m^2\right)\Phi=\frac{1}{\sqrt{-g}}\partial_{\mu}\left(\sqrt{-g}g^{\mu\nu}\partial_{\nu}\Phi\right)+m^2 \Phi=0,
    \label{KG}
\end{equation}
  with $\sqrt{-g}=\sin{\beta}\left(\dfrac{r_{AdS}^2}{\cos{\beta}}\right)^{4}$ and
  \begin{equation}
  g_{\mu\nu}=\left(\begin{array}{cccc}
      \dfrac{r_{AdS}^2}{\cos^{2}{\beta}}r^2 & -\dfrac{r_{AdS}^2}{\cos^{2}{\beta}} & 0 & 0  \\
      -\dfrac{r_{AdS}^2}{\cos^{2}{\beta}} & 0 & 0 & 0\\
      0 & 0 & -\dfrac{r_{AdS}^2}{\cos^{2}{\beta}} & 0\\
      0 & 0 & 0 & -r_{AdS}^2 \tan^{2}{\beta}
  \end{array}\right),
  \label{g1}
  \end{equation}

\begin{equation}
  g^{\mu\nu}=\left(\begin{array}{cccc}
      0 & -\dfrac{\cos^{2}{\beta}}{r_{AdS}^2} & 0 & 0  \\
      -\dfrac{\cos^{2}{\beta}}{r_{AdS}^2} & -\dfrac{\cos^{2}{\beta}}{r_{AdS}^2}r^2 & 0 & 0\\
      0 & 0 & -\dfrac{\cos^{2}{\beta}}{r_{AdS}^2} & 0\\
      0 & 0 & 0 & -\dfrac{\cot^{2}{\beta}}{r_{AdS}^{2}}
  \end{array}\right),
  \label{g2}
  \end{equation}
 where $g^{\mu\nu}$ is the inverse metric of $g_{\mu\nu}$. Plugging Eqs.(\ref{g1}) and (\ref{g2}) into Eq.(\ref{KG}) and upon some algebraic manipulations, one finds
 \begin{eqnarray}
     \nonumber 0&=&-2\frac{\sin{\beta}}{\cos^{2}{\beta}}\partial_{v}\partial_{r}\Phi -\frac{\sin{\beta}}{\cos^{2}{\beta}}\partial_{r}\left(r^{2}\partial_{r}\Phi\right)-\partial_{\beta}\left(\frac{\sin{\beta}}{\cos^{2}\beta}\partial_{\beta}\Phi\right)-\frac{1}{\sin{\beta}\cos^{2}{\beta}}\partial^{2}_{\theta}\Phi+\\
     &+&m^2 \frac{\sin{\beta}}{\cos^{4}{\beta}}\Phi.
     \label{KG1}
 \end{eqnarray}
In order to solve it, we decompose the scalar field in modes as shown in the ansatz below:
\begin{equation}
    \Phi=\sum\limits_{l,m^{\prime}}\Psi_{m^{\prime}}(v,r)Y_{l}(\theta)\phi_{l,m^{\prime}}(\beta),
    \label{exp}
\end{equation}
where $l,m^{\prime}$ are quantum numbers featuring the general solution and the spherical harmonic function $Y_{l}$ satisfies
\begin{equation}
    \Box_{S^{1}}Y_{l}(\theta)=-l^{2}Y_{l}(\theta)
    \end{equation}
where $ \Box_{S^{1}}=\partial^{2}_{\theta}$. By substituting the expansion in modes (\ref{exp}), Eq.(\ref{KG1}) can be set into the form
\begin{eqnarray}
   \nonumber 0&=& \frac{1}{\Psi(v,r)}\left[2\partial_{v}\partial_{r}\Psi(v,r)+\partial_{r}\left(r^2 \partial_{r}\Psi(v,r)\right)\right]+\frac{\cos^{2}\beta}{\sin\beta}\frac{1}{\phi(\beta)}\partial_{\beta}\left(\frac{\sin\beta}{\cos^{2}\beta}\partial_{\beta}\phi({\beta})\right)-\\
   &-&\frac{l^{2}}{\sin^{2}\beta}-\frac{m^2}{\cos^{2}\beta}.
   \label{kjy}
\end{eqnarray}
The labels $l$ and $m^{\prime}$ were omitted above for simplicity. It is noteworthy that Eq.(\ref{kjy}) might be conveniently rewritten as two differential equations by introducing a separation constant ($\xi^2)$, namely, 
\begin{eqnarray}
   \label{11} 0&=&2\partial_{v}\partial_{r}\Psi(v,r)+\partial_{r}\left(r^2 \partial_{r}\Psi(v,r)\right)-\xi^2 \Psi(v,r);\\
    \label{22}0&=&\partial_{\beta}\left(\frac{\sin\beta}{\cos^{2}\beta}\partial_{\beta}\phi(\beta)\right)-\left(\frac{l^{2}}{\sin\beta \cos^{2}\beta}+\frac{m^2 \sin\beta}{\cos^{4}\beta}-\xi^2 \frac{\sin\beta}{\cos^2 \beta}\right)\phi(\beta).
\end{eqnarray}
Eq.(\ref{11}) is the partial differential equation (PDE) related to the $AdS_{2}$-section of the whole space ($AdS_{4}$) whilst Eq.(\ref{22}) is the ordinary differential equation concerned with the radial function, $\phi(\beta)$. We shall deal only with the latter since the former has already been discussed in the literature \cite{Lucietti, Cvetic}.


Our goal now is to provide an analytical solution for the radial equation (\ref{22}). To begin with, it is convenient to introduce the new coordinate $y=\sin^{2}\beta$ in order to use the Sturm-Liouville theory \cite{Zettl}. By doing so, the radial equation looks like
\begin{eqnarray}
    L[\phi(y)]=\omega^2 r(y)\phi(y),
    \label{ews}
\end{eqnarray}
where $L$ is the Sturm-Liouville operator. More explicitly,
\begin{eqnarray}
    L[\phi(y)]&=&-\partial_{y}\left(p(y)\partial_{y}\phi(y)\right)+q(y)\phi(y);\\
        p(y)&=&\frac{y}{(1-y)^{1/2}};\\
    q(y)&=&\frac{m^2}{4(1-y)^{5/2}}+\frac{l^2}{4}\frac{1}{y(1-y)^{3/2}};\label{qy}\\
    r(y)&=&\frac{1}{(1-y)^{3/2}};\\
    \omega^2 &=&\frac{\xi^2}{4}.
\end{eqnarray}
Note that $y=0$ ($\beta=0$) and $y=1$ ($\beta=\frac{\pi}{2}$) are singular points. Furthermore, the function $q(y)$ should be interpreted as an effective potential; so that now the ``mass'' term contains curvature contributions (the second term in the \textit{r.h.s} of Eq.(\ref{qy})) arising from the dimensional reduction procedure, apart from the usual squared mass $m^2$. The following general boundary conditions should be fulfilled 
\begin{eqnarray}
    \alpha_{1}\phi(0)+\alpha_{2}\phi^{\prime}(0)=0, \,\,\,\, \gamma_{1}\phi(1)+\gamma_{2}\phi^{\prime}(1)=0,
\end{eqnarray}
with $\alpha$'s and $\gamma$'s arbitrary constants, and the prime stands for derivative with respect to $y$. Since we are interested in well-defined solutions at the singular points it makes necessary imposing regularity conditions. This is pursued by investigating the behavior of the solutions around each of two singular points ($y=0$, origin, and $y=1$, boundary). 

Proceeding with the redefinition
\begin{equation}
    \phi(y)=(1-y)^{\lambda}y^{\sigma}\varphi(y),
\end{equation}
to facilitate the algebraic manipulations, Eq.(\ref{ews}) reduces to the hypergeometric differential equation \cite{Abr}
\begin{eqnarray}
   \nonumber 0&=&y(1-y)\varphi^{\prime\prime}+\left(\left(-2\sigma-2\lambda-\frac{1}{2}\right)y+2\sigma+1\right)\varphi^{\prime}+\\
    &+&\left(-\left(\sigma+\lambda\right)^2 +\frac{1}{2}(\lambda+\sigma)+\frac{\xi^2}{4}\right)\varphi,
    \label{kk}
\end{eqnarray}
where the coefficients must be identified as 
\begin{eqnarray}
    4\sigma^{2} &=&l^2;\\
    4\lambda^2 -6\lambda&=&m^2,
\end{eqnarray}
whose solutions are
\begin{eqnarray}
    \sigma_{\pm} &=&\pm \frac{l}{2};\\
    \lambda_{\pm}&=&\frac{3}{4}\pm\frac{1}{4}\sqrt{9+4m^2}.
    \label{lm}
\end{eqnarray}
Recalling that the mass should satisfy the  Breitenlohner-Freedman (BF) bound \cite{Breit} 
\begin{equation}
m^2 \ge -\frac{9}{4},
\label{BFb}
\end{equation}
to guarantee real solutions. Moreover, there are two linear independent hypergeometric solutions for Eq.(\ref{kk}) depending on the parameter $\sigma$ and $\lambda$. It is straightforward to check the leading behavior of the solution near the single points. Around the origin ($y=0$), the function $\phi(y)$ falls off as $ y^{\sigma_{\pm}}$ while, near the boundary ($y=1$), $\phi(y)$ falls off as $(1-y)^{\lambda_{\pm}}$. To shed more light on this,  we fix $\lambda=\lambda_{+}$ without loss of generality since the asymptotic behavior at the origin depends only on $\sigma_{\pm}$. In this case, using the general formulas of \cite{Math} and restoring the coordinate $\beta$, we are able to find the full analytical solutions of Eq. (\ref{ews}) near the origin $\beta=0$, namely,
\begin{eqnarray}
\label{ks}\phi_{1}(\beta)&=&\sin^{l}(\beta)\cos^{2\lambda_{+}}(\beta)\, {}_2 F_1 \left(a_{+},b_{+};l+1;\sin^{2}(\beta)\right);\\
\nonumber\phi_{2}(\beta)&=&\phi_{1}(\beta)\ln{\left(\sin^{2}(\beta)\right)}-\bigg[\sum_{k=1}^{l}\frac{l!(k-1)!}{(l-k)!(1-a_{+})_{k}(1-b_{+})_{k}}\left(-\sin^{2k}(\beta)\right)+\\
\label{ks1}&+&\sum_{k=0}^{\infty}\frac{(a_{+})_{k}(b_{+})_{k}}{(l+1)_{k} k!}g_{k}\sin^{2k}(\beta)\bigg],
\end{eqnarray}
where 
\begin{equation}
    \begin{split}
        a_{+}&=\lambda_{+}-\frac{1}{4}\left(1-\sqrt{1+4\xi^2}\right)+\frac{l}{2};\\
        b_{+}&=\lambda_{+}-\frac{1}
        {4}\left(1+\sqrt{1+4\xi^2}\right)+\frac{l}{2},
        \end{split}
        \end{equation}
    and the other quantities are defined by
    \begin{equation}
    \begin{split}
        \psi(x)&=\frac{\Gamma^{\prime}(x)}{\Gamma(x)};\\
        g_{k}&=\psi(a_{+}+k)+\psi(b_{+}+k)-\psi(1+k)-\psi(l+1+k);\\
        (a_{+})_{k}&=\frac{\Gamma(a_{+}+k)}{\Gamma(k)},
    \end{split}
\end{equation}
similar to \cite{Math}.

As a result of the AdS/CFT property \cite{Witten, Kraus}, the boundary action must vanish by imposing regularity conditions at the origin $y=0$. In other words, 
\begin{equation}
   S_{boundary}=\int d^{3}x\sqrt{-h}\,g^{\beta\beta}\Phi\partial_{\beta}\Phi\big{|}_{\beta=0}=0,
\end{equation}
which just holds for the solution $\phi_{1}(\beta)$ while  for $\phi_{2}(\beta)$ does not. Hence, $\phi_{2}(\beta)$ is an unacceptable solution.

We now turn our attention to the solutions near the boundary $y=1\, (\beta=\pi/2)$. In this case, it is convenient to rewrite the radial equation in terms of the coordinate $y^{'}=\cos^{2}\beta$ or  $y^{'}=1-y$. Note however that, by using the linear properties of hypergeometric functions \cite{Abr}, one recovers a similar equation to Eq.(\ref{ews}) with the difference that $y$ should be replaced by $(1-y)$. In this scenario, the general solution looks like
\begin{equation}
    \phi(\beta)=A\phi_{3}(\beta)+B\phi_{4}(\beta),
\end{equation}
where $A$ and $B$ are arbitrary constants and $\phi_{3}(\beta)$ and $\phi_{4}(\beta)$ are linear independent solutions near the boundary. As it was discussed before, the asymptotic behavior of $\phi(\beta)$ falls off as $\cos^{2\lambda_{\pm}}(\beta)$ near the boundary, thereby we can pick $\sigma=\sigma_{+}$ without loss of generality. Moreover, it is worth mentioning that the explicit solutions will depend on the difference between $\lambda_{+}$ and $\lambda_{-}$, and here we define it by $\nu=(\lambda_{+}-\lambda_{-})$; thus, by using Eq. (\ref{lm}), it is easy to see that
\begin{equation}
    \nu=\frac{1}{2}\sqrt{9+4m^2}.
\end{equation}
As a result of the BF bound (\ref{BFb}), we have $\nu\ge 0$, which ensures stable solutions. In this way, the general solution can be split into three different cases depending on $\nu$:

\begin{enumerate}
    \item The $\nu$ is a non-integer number. In this case,
    \begin{eqnarray}
        \nonumber\phi(\beta)&=&\frac{\Gamma(l+1)\Gamma(-\nu)}{\Gamma(a_{-})\Gamma(b_{-})}\sin^{l}(\beta)\cos^{2\lambda_{+}}(\beta)\,{}_2 F_1 \left(a_{+},b_{+};(1+\nu);\cos^{2}(\beta)\right)+\\
        &+&\frac{\Gamma(l+1)\Gamma(\nu)}{\Gamma(a_{+})\Gamma(b_{+})}\sin^{l}(\beta)\cos^{2\lambda_{-}}(\beta)\,{}_2 F_1 \left(a_{-},b_{-};(1-\nu);\cos^{2}(\beta)\right),
    \end{eqnarray}
    where 
    \begin{eqnarray}
        a_{-}&=&\lambda_{-}-\frac{1}{4}\left(1-\sqrt{1+4\xi^2}\right)+\frac{l}{2};\\
        b_{-}&=&\lambda_{-}-\frac{1}
        {4}\left(1+\sqrt{1+4\xi^2}\right)+\frac{l}{2};\\
        A&=&\frac{\Gamma(l+1)\Gamma(-\nu)}{\Gamma(a_{-})\Gamma(b_{-})};\\
        B&=&\frac{\Gamma(l+1)\Gamma(\nu)}{\Gamma(a_{+})\Gamma(b_{+})}.
    \end{eqnarray}
    \item The $\nu$ is an integer number. In this situation, the general solution is given by
    \begin{eqnarray}
\nonumber\phi(\beta)&=&\sin^{l}(\beta)\cos^{2\lambda_{-}}(\beta)\frac{\Gamma(l+1)\Gamma(\nu)}{\Gamma(a_{+})\Gamma(b_{+})}\sum_{k=0}^{\nu-1}\left(\frac{(a_{+}-\nu)_{k} (b_{+}-\nu)_{k}}{k!(1-\nu)_{k}}\cos^{2k}(\beta)\right)-\\
\nonumber &-&\sin^{l}(\beta)\cos^{2\lambda_{+}}(\beta)\frac{(-1)^{\nu}\Gamma(l+1)\Gamma(\nu)}{\Gamma(a_{+} - \nu)\Gamma(b_{+}-\nu)}\sum_{k=0}^{\infty}\frac{(a_{+})_{k} (b_{+})_{k}}{k!(k+\nu)!}\cos^{2k}(\beta)\times\\&\nonumber\times& \big[\ln{(\cos^{2}(\beta))}-\\
&-&\psi(k+1)-\psi(k+\nu+1)+\psi(a_{+}+k)+\psi(b_{+}+k)\big],
    \end{eqnarray}
    where 
    \begin{eqnarray}
        A&=&\frac{(-1)^{\nu}\Gamma(l+1)\Gamma(\nu)}{\Gamma(a_{+} - \nu)\Gamma(b_{+}-\nu)};\\
        B&=&\frac{\Gamma(l+1)\Gamma(\nu)}{\Gamma(a_{+})\Gamma(b_{+})}.
    \end{eqnarray}
    \item The case $\nu=0$. The solution reduces to 
    \begin{eqnarray}
        \nonumber\phi(\beta)&=& \sin^{l}(\beta)\cos^{3/2}(\beta)\bigg\{\frac{\Gamma(l+1)}{\Gamma(a_{+})\Gamma(b_{+})}\sum_{k=0}^{\infty}\frac{(a_{+})(b_{+})}{(k!)^{2}}\bigg[2\psi(k+1)-\psi(a_{+}+k)-\\
        &-&\psi(b_{+}+k)-\ln{(\cos^{2}(\beta))}\bigg]\cos^{2k}(\beta)\bigg\}
    \end{eqnarray}
    with
    \begin{eqnarray}
        A=B=\frac{\Gamma(l+1)}{\Gamma(a_{+})\Gamma(b_{+})}.
    \end{eqnarray}
\end{enumerate}

Now, it is important to investigate  the behavior of the general solution near the boundary. For the first case ($\nu$ is a non-integer number), it is straightforward to check from the properties of the hypergeometric functions that $\phi(\beta)$ falls off as $\cos^{2\lambda_{\pm}}(\beta)$ near the boundary. The fluctuations modes corresponding to $\lambda_{-}$ fail to be square integrable, as a consequence, they are classified as non-normalizable. On the other hand, the modes associated with $\lambda_{+}$ are normalizable, as a result, they are classified as normalizable modes. Therefore, near the boundary, the asymptotic behavior of the general solution can be cast into the following compact form
\begin{equation}
    \Phi(Z,\beta) \sim \Phi_{1}(Z)\cos^{2\lambda_{-}}(\beta)+\,\Phi_{2}(Z)\cos^{2\lambda_{+}}(\beta),
\end{equation}
where $Z=(v,r,\theta)$ is a shorthand notation to describe the three-dimensional boundary coordinates  and $\Phi_{1}(Z)$ and $\Phi_{2}(Z)$ are arbitrary fields living in the boundary.

For our purpose in this work, we must only select the normalizable modes since it is well known that the non-normalizable modes lead to a non-trivial contribution to the boundary action \cite{Witten2}. For $\nu>1$, this is achieved by imposing the vanishing of the reciprocal gamma functions of the non-normalizable modes, such requirement leads to the following quantization condition on the separation constant
\begin{equation}
\xi=\pm\sqrt{\left(2\lambda_{+}+l+2n\right)\left(2\lambda_{+}+l+2n-1\right)},\,\,\, \mbox{with}\,\,\, n=0,1,2...
\label{qc}
\end{equation}
In such case, the coefficient $B$ vanishes for both cases: when $\nu$ is an integer number and a non-integer number. Other than for $0<\nu<1$ in which both modes are normalizable. In this case, the mode proportional to $\cos^{2\lambda_{-}}(\beta)$ is dominant near the boundary, such mode is selected by the following quantization condition
\begin{equation}
\xi=\pm\sqrt{\left(2\lambda_{-}+l+2n\right)\left(2\lambda_{-}+l+2n-1\right)},\,\,\, \mbox{with}\,\,\, n=0,1,2...
\label{qc1}
\end{equation}
When $\nu=0$, both modes coincide, $\lambda_{-}=\lambda_{+}=3/4$, and are normalizable.

\subsection{Presence of a self-interacting scalar field}
\label{subsection}
In this subsection, unlike the previous one, we allow for the backreaction of the scalar field on the background geometry. In other words, the energy-momentum tensor of the scalar field is not neglected in the gravitational field equations. Furthermore, we also include scalar field self-interactions through the scalar potential $V(\Phi)$. Explicitly, the action  of the scalar field is now given by
\begin{equation}
    S[\Phi]=-\frac{1}{2}\int d^{\mathcal{D}}x\sqrt{-g}\bigg(g^{\mu\nu}\partial_{\mu}\Phi\partial_{\nu}\Phi-2V(\Phi)\bigg),
    \label{sfe}
\end{equation}
leading to the equation of motion for $\Phi$ field in the form $\Box \Phi + V_\Phi = 0$, with $V_\Phi=dV/d\Phi$. 

In this new environment, we want to search for the presence of brane-like configurations having a thickness due to the self-interacting scalar, as it happens in the standard $AdS_5$ case with a single extra dimension of infinite extent \cite{SB1,SB11,SB2,SB3}.
Towards this goal, let us consider the scalar field $\Phi$ depending only on the extra dimension $w$. The non-vanishing components of the energy-momentum tensor are then
\begin{equation}    T^{a}_{b}=\delta^{a}_{b}\left(\frac{1}{2}\Phi^{\prime 2}+V(\Phi)\right);\,\,\, T^{D}_{D}= -\frac{1}{2}\Phi^{\prime 2}+V(\Phi).
\end{equation}
Substituting these results in the field equations (\ref{kj}) and proceeding with the regularization, $\mathcal{D}\rightarrow 4$, we are able to find 
\begin{eqnarray}
   \label{oneE} \kappa^2 \left(\frac{1}{2}\Phi^{\prime 2}+V(\Phi)\right)&=&-3(A^{\prime})^2 -2 A^{\prime\prime}+2\alpha\kappa^2 \left(3(A^{\prime})^4 + 4A^{\prime\prime}(A^{\prime})^2 \right),\\
\label{twoE}\kappa^2 \left(-\frac{1}{2}\Phi^{\prime 2}+V(\Phi)\right)&=&-3(A^{\prime})^2+6\alpha\kappa^2 (A^{\prime})^4,
\end{eqnarray}
or yet
\begin{eqnarray}
   \label{einstein1} V(\Phi)&=& - \frac1{\kappa^2} A^{\prime\prime} 
   -\frac3{\kappa^2} A^{\prime2}  + 6 \alpha  A^{\prime4} +  4 \alpha A^{\prime2} A^{\prime\prime},
   \\ \label{einstein2}
\Phi^{\prime2} &=& - \frac2{\kappa^2} A^{\prime\prime} + 8 \alpha A^{\prime2} A^{\prime\prime}. 
\end{eqnarray}
Using the metric \eqref{metric}, the  equation of motion for $\Phi$ becomes
\be
\Phi^{\prime\prime} + 3 A^\prime \phi^\prime + V_\Phi = 0. 
\ee

 In order to get a first-order framework, we follow \cite{Gbrane} and introduce an arbitrary function of the scalar field $W(\Phi)$ such that
\begin{equation}\label{fo1}
A^{\prime} =-\frac13\; {W(\Phi)}.
\end{equation}
We use this in Eq. \eqref{einstein2} to get 
\begin{equation}
\label{fo2} 
\Phi^\prime = \frac2{3\kappa^2} W_\Phi  - \frac{8\alpha}{27} W(\Phi)^2  W_\Phi,
\end{equation}
where $W_\Phi=dW/d\Phi$. We see that the term dependent on $\alpha$ brings the possibility of new types of solutions.
We substitute the equations Eqs. \eqref{fo1} and \eqref{fo2} in Eq. \eqref{einstein1} to obtain the potential
\be\label{potW}
V(\Phi) = \frac2{9\kappa^4} W_\Phi^2  - \frac{1}{3\kappa^2} W(\Phi)^2
+\frac{2\alpha}{81\kappa^2} W(\Phi)^2\left(3 W(\Phi)^2 \kappa^2 - 8 W_\Phi^2 \right) + \frac{32\alpha^2}{729}W(\Phi)^4 W_\Phi^2.
\ee

There are homogeneous solutions of the first order equation given by Eq.~\eqref{fo2}. We can in particular consider the two possibilities, obeying
\begin{subequations}
\begin{eqnarray}
\label{way1} W_\Phi &=& 0, \\
\label{way2} W^2(\Phi) &=&\frac{9}{4\alpha \kappa^2}. 
\label{Min2}
\end{eqnarray}
\end{subequations}
In the first case, solutions $\bar\Phi$ that obey the relation \eqref{way1} lead to the potential with the following behavior 
\begin{subequations}
\begin{eqnarray}
\label{V01}
V(\bar\Phi) &=&   - \frac{1}{3\kappa^2} W(\bar\Phi)^2
+\frac{2\alpha}{27} W(\bar\Phi)^4, \\
V_\Phi(\bar \Phi) &=& 0, \\
V_{\Phi\Phi} (\bar \Phi) &=& \frac{2}{729\kappa^4}W_{\Phi\Phi}(4\alpha \kappa^2 W^2 - 9)(2(4\alpha \kappa^2 W^2-9)W_{\Phi\Phi} + 27 W \kappa^2 ).
\label{V02}
\end{eqnarray}
\end{subequations}
The critical values have characteristics that may depend on the GB contribution. Moreover, in the second case the system only makes sense in the presence of the Gauss term $(\alpha \neq 0)$. Here one observes that the potential has the following features
\begin{subequations}
\begin{eqnarray}
V(\bar\Phi) &=&   -\frac{3}{8\kappa^4 \alpha}, \\
V_\Phi(\bar \Phi) &=& 0, \\
V_{\Phi\Phi}(\bar \Phi) &=& \frac{4W_\Phi^2}{81\kappa^2}(27+16 \alpha W_\Phi^2), 
\end{eqnarray}
\end{subequations}
in this case always identifying minima that lead to negative values of the potential. 

The above general results motivate us to further explore specific models, investigating how adding new gravitational term changes the standard scenario. Let us now study two specific situations, controlled via the function $W$, with
\begin{subequations}
\begin{eqnarray}
 \label{W1}
W_I(\Phi)&=& a\,\Phi,  \\
W_{II}(\Phi)&=&a \sin(b\Phi).
 \label{W2}
 \end{eqnarray} 
 \end{subequations}
These functions correspond to {\it{type-I}} and {\it{type-II}} models, where $a$ and $b$ are real parameters.

\subsubsection{{Type-I} model}
\begin{figure}[ht]
    \centering
\includegraphics[width=0.40\textwidth]{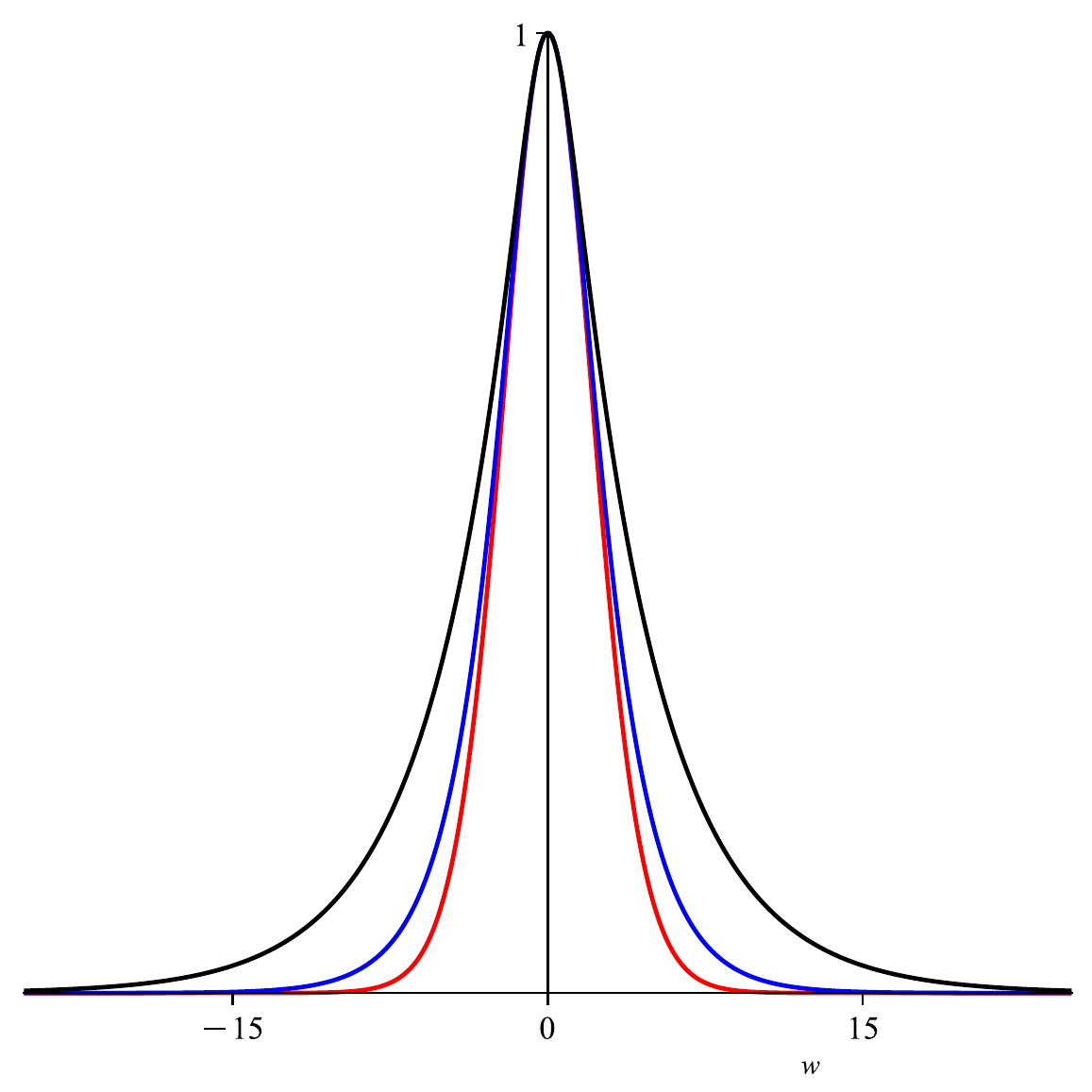}
    \caption{The warp factor depicted with $A(w)$ in Eq. \eqref{A1}, for $a=1$, $\kappa=1$ and $\alpha=1/4$ (red), $1$ (blue), and $4$ (black).}
    \label{figWarp1}
\end{figure}

The first example that we consider is described by the simplest choice for $W(\Phi)$, the linear function. Using  Eq.~\eqref{W1}, the potential \eqref{potW} is written in the form 
 \be \label{pot1}
V(\Phi)= \frac{2a^2}{9\kappa^4} - \frac{a^2(16a^2\alpha+27)}{81\kappa^2}\Phi^2
+\frac{2a^4 \alpha (16a^2\alpha +27)}{729}\Phi^4.
\ee
The first order equation given by Eq.~\eqref{fo1} becomes
\be
 \label{dpEX1}
\Phi^\prime = \frac{8a^3 \alpha}{27}\left( \frac{9}{4a^2  \kappa^2 \alpha} - \Phi^2\right).
\ee
Using the Eq.~\eqref{Min2}, we find the asymptotic limits of the solution at $\Phi_\pm = \pm 3/2\kappa a\sqrt{\alpha}$, which are the minima of the potential, whose values of $V_{\Phi\Phi}$ are given by Eq.~\eqref{V02}. We can see from the potential in Eq. \eqref{pot1}, for instance, that it engenders spontaneous symmetry breaking, so it support kink-like solutions. In order to explicitly describe this case, we go on and solve Eq.~\eqref{dpEX1}, leading to the kink solution   
\be 
\Phi(w)=\frac{3}{2 \kappa a \sqrt{\alpha}} \tanh \left(\frac{4\sqrt{\alpha} a \, w}{9 \kappa} \right),
\ee
where we have taken the positive sign and chosen the integration constant such that $\phi(0)=0$. Knowing the field profile, we can rewrite equation \eqref{fo1} as
\be
A^\prime = -\frac{1}{2 \kappa \sqrt{\alpha}} \tanh \left(\frac{4\sqrt{\alpha} a \, w}{9 \kappa} \right),
\ee
which gives the warp function
\be \label{A1}
A(w)=\frac{9}{8a^2 \alpha} \ln \left(\sech \left(\frac{4\sqrt{\alpha} a \, w}{9 \kappa} \right) \right),
\ee
with $A(0)=0$. As we can see, this model leads to analytical solutions for both the scalar field and the warp factor. Interestingly, we notice that for $|w|>>1$ one gets
\be
\lim_{w \to \pm \infty} A(w) =  -\dfrac{1}{\sqrt{2\alpha}\kappa}|w|.
\ee
As expected, we have obtained the same result shown in Eq. \eqref{A_w}. Furthermore, using the metric \eqref{metric},  which is here controlled by $A(w)$ in \eqref{A1}, we see that the warp factor $e^{2A}$ has the adequate profile, with its thickness increasing as we increase the parameter $\alpha$ that responds for the modification introduced by the GB contribution. This is illustrated in Fig. \ref{figWarp1}.

\subsubsection{Type-II model}

Now, look at the second model presented in equation \eqref{W2}, which is an extension of the sine-Gordon model. In this case, using \eqref{potW}, \eqref{dpEX1} and \eqref{fo1}, we obtain
\begin{subequations}
\begin{eqnarray}
\nonumber \label{Vsin}
V(\Phi) = &&-\frac{(9-2\alpha a^2 \kappa^2)a^2}{27\kappa^2} + \frac{(9-4\alpha a^2 \kappa^2)(27\kappa^2 + 18b^2 - 8 a^2 b^2 \alpha \kappa^2 )a^2}{729\kappa^4} \cos^2(b\Phi) \\
&&+\frac{2\alpha a^4 (27\kappa^2 + 72b^2 - 32 a^2 \alpha b^2 \kappa^2)}{729\kappa^2} \cos^4(b\Phi)
+\frac{32a^6 \alpha^2 b^2}{729}\cos^6(b\Phi),
\\
\label{fo1Ex2}\Phi^\prime =&& \frac{2ab}{3\kappa^2} \cos(b\Phi) \left(1 - \frac{4a^2 \alpha \kappa^2 }{9} \sin^2(b\Phi)\right), \\  
A^\prime =&& -\frac{a}{3} \sin(b\Phi). \label{fo2Ex2}
\end{eqnarray}
\end{subequations}
When $\alpha=0$, we get  $V(\Phi) = -a^2/3\kappa^2 + (3\kappa^2 + 2b^2 )a^2/9\kappa^4 \cos^2(b\Phi)$ and its first order equation  $\Phi^\prime = {2ab}/{3\kappa^2} \cos(b\Phi)$, as in the original sine-Gordon model. An interesting case is when $\alpha=\alpha_1$, where $\alpha_1=9/4a^2\kappa^2$. Under this condition, Eqs. \eqref{Vsin} and \eqref{fo1Ex2} give
\begin{subequations}
\begin{eqnarray}
\Phi^\prime &=& \frac{2ab}{3\kappa^2} \cos^3 (b\Phi),\\
V(\Phi)&=&-\frac{a^2}{6\kappa^2} +\frac{a^2}{6\kappa^2}\cos^4(b\Phi) 
+\frac{2a^2b^2}{9\kappa^4}\cos^6(b\Phi).\label{poten}
\end{eqnarray}
\end{subequations}
To determine the asymptotic values of the solutions of Eq.~\eqref{fo1Ex2}, we can use the two ways to make it vanish. By applying \eqref{way1} and  
\eqref{way2}, respectively, we get 
\begin{subequations}
    \begin{eqnarray}
 \label{minn}\Phi_n &=& \frac{(2n+1)\pi}{2b}, \\
\label{minm}\Phi_{m_\pm} &=& \frac{m\pi}b \pm \frac{1}b {\arcsin\left(\frac{3}{2\kappa a\sqrt{\alpha}}\right)}, 
\end{eqnarray} 
\end{subequations}
where $n$ and $m$ are integer numbers. For the $\Phi_n$ case, we get
\begin{eqnarray}
V_{\Phi\Phi}(\Phi_n)&=& \frac{2a^2b^2(4\alpha a^2 \kappa^2 - 9)(8a^2 b^2 \alpha \kappa^2 - 18b^2 - 27\kappa^2)}{729\kappa^4} \\
&=& \frac{9(\alpha - \alpha_1)(\alpha - \alpha_{2})}{4\alpha_1\kappa^2(\alpha_1 - \alpha_{2})^2},
\end{eqnarray}
with $\alpha_2=\alpha_1(1+(3\kappa^2/2b^2))$. From the above expression, we can determine that $\Phi_n$ represents the local maxima of the potential for $\alpha_1<\alpha<\alpha_2$. In other regions of parameter values, $\Phi_n$ corresponds to the minima of the potential. Furthermore, the existence of $\Phi_{m_\pm}$ is limited to values of $\alpha$ greater than or equal to $\alpha_1$, and they always represent minima of the potential. In the special case of $\alpha=\alpha_1$, these minima collapse to the values given by Eq.~\eqref{minn}.

\begin{figure}[ht]
    \centering    \includegraphics[width=0.60\textwidth]{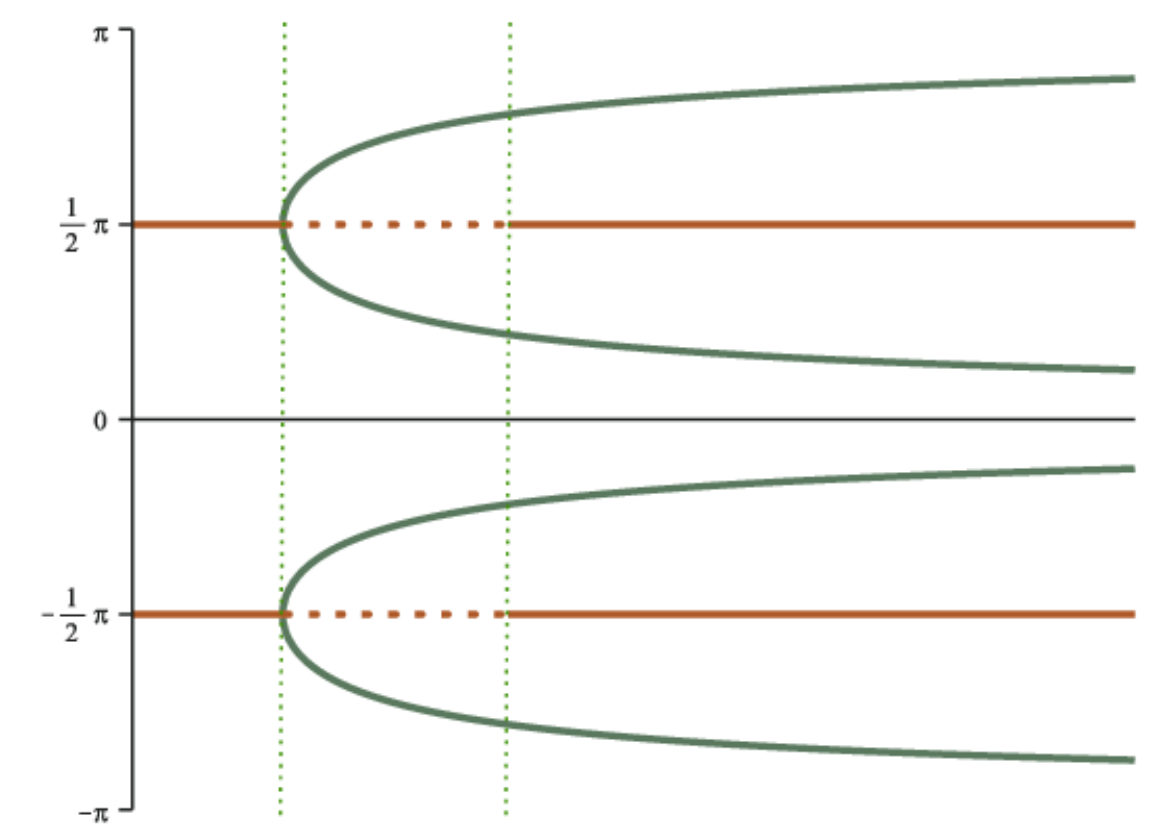}
    \caption{The structure of the critical points $\Phi_n$ (brown) and $\Phi_{m_\pm}$(green) associated with the potential  $V(\Phi)$ in Eq.~\eqref{Vsin} as a function of $\alpha$. The solid (dotted) lines represent values of minimum (maximum).}
    \label{fig2}
\end{figure}

\begin{figure}[ht]
    \centering
    
\includegraphics[width=0.60\textwidth]{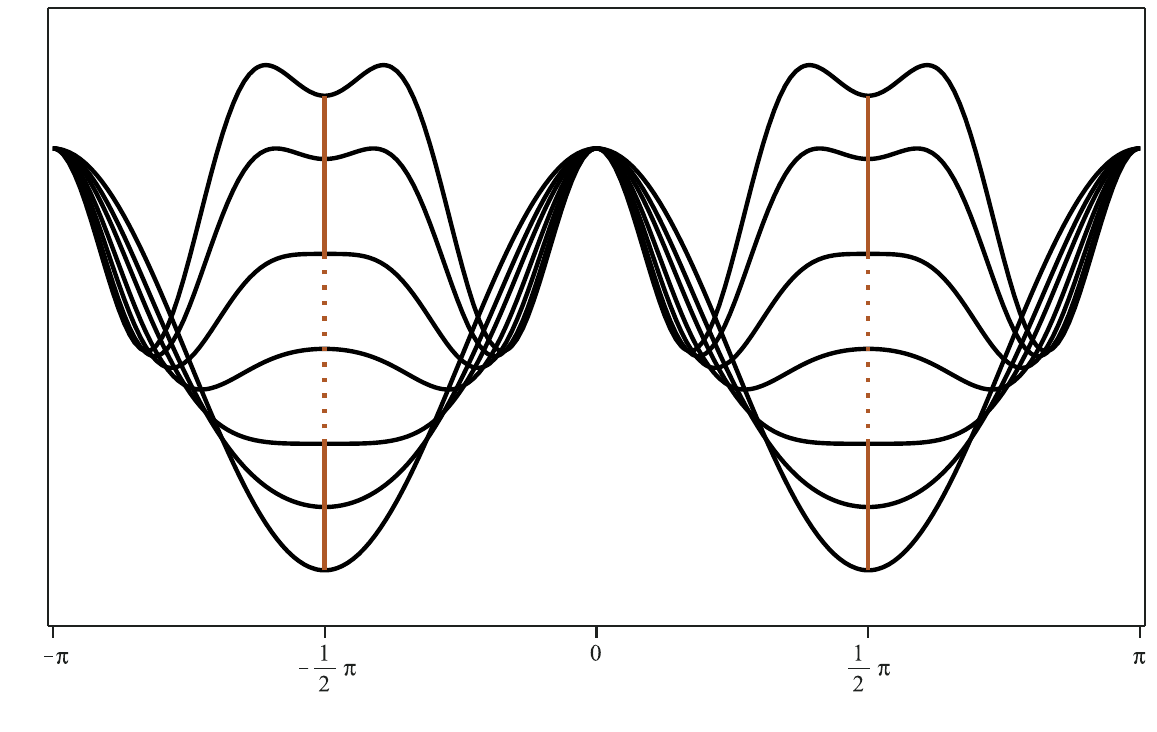}
    \caption{The potential $V(\phi)$ given by Eq. \eqref{poten} with $\kappa=1$, $b=1$, $a=1$ and some values of $\alpha$ parameter.  The vertical brown line segments represent the regions where the critical points $\Phi_n$ are minimum (solid line) and maximum (dotted line).}
    \label{fig3}
\end{figure}

Considering the conditions obtained above, we display in Fig. \ref{fig2}, the critical points in terms of $\alpha$, the green and brown curves representing $\Phi_{m_\pm}$ and  $\Phi_n$, respectively. Minima and maxima are represented by solid and dotted lines, respectively. In the Fig. \ref{fig3}, we display the potential $V(\phi)$ for some values for $\alpha$. The vertical brown lines denote regions where $\Phi_n$ are minima (solid lines) or maxima  (dotted lines). 

\begin{figure}[ht]
    \centering
\includegraphics[width=0.40\textwidth]{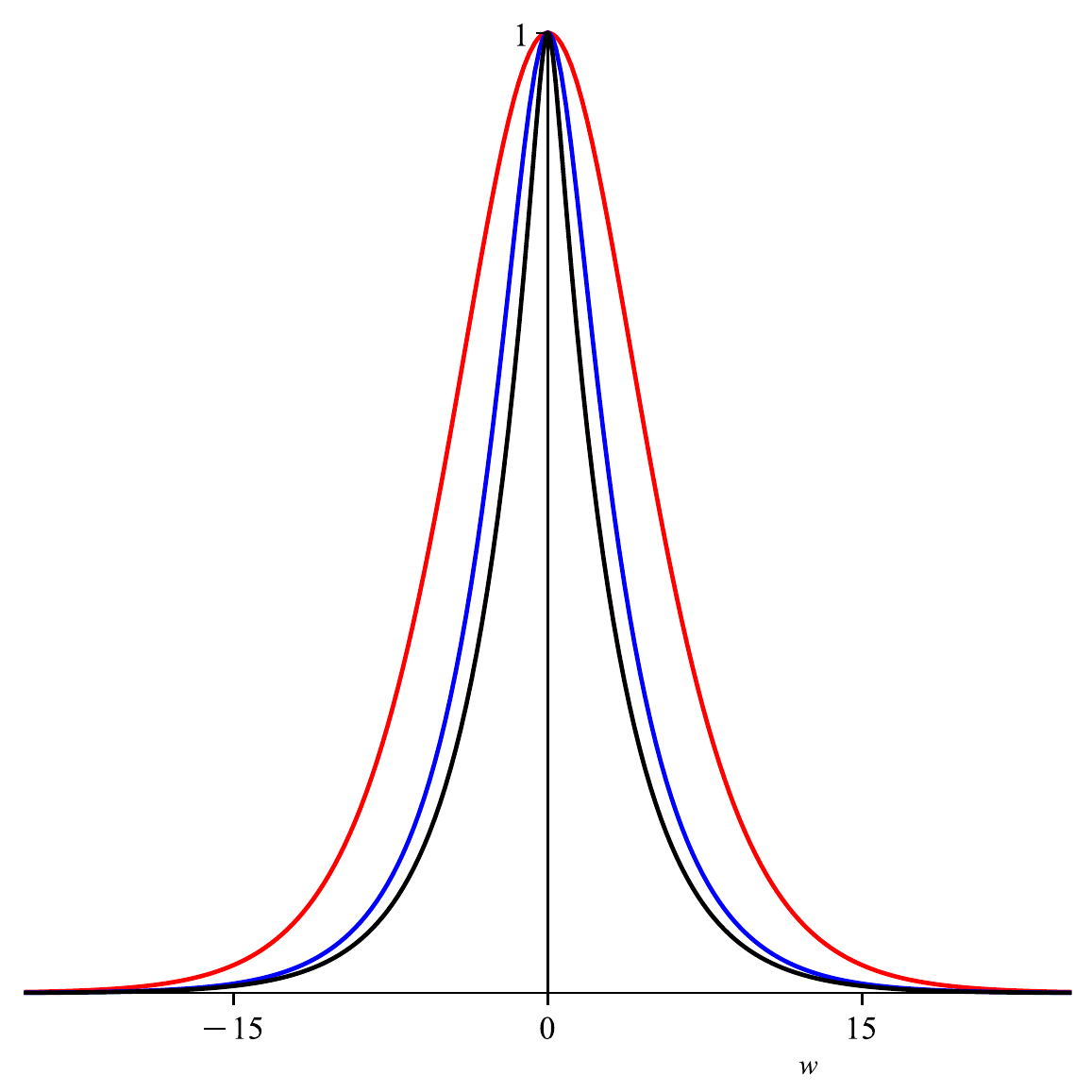}
    \caption{The warp factor depicted with $A(w)$ in Eq. \eqref{A20} for $\alpha=\alpha_1/4.$ We take $a=1$, $\kappa=1$ and $b=1/2$ (red), $1$ (blue), and $2$ (black).}
    \label{figWarp21}
\end{figure}

We can write the scalar field as 
\be 
\Phi=\frac{m \pi }{b} \pm \frac{\arcsin(\chi)}b,
\ee 
with $m$ integer. With this, we rewrite the Eq.\eqref{fo1Ex2} as
\be
\frac{d\chi}{dw} =\pm \frac{8a^3b^2}{27} (1-\chi^2)
\left(\frac{9}{4a^2\alpha \kappa^2}- \chi^2\right).
\ee 
This equation can be solved numerically, but it is not possible to find analytical solutions for the equation for arbitrary choices of parameters. However, using results from Ref. \cite{Bazeia:2006pj}, we can obtain analytical solutions for two specific situations. For $\alpha=\alpha_1/4= 9/16a^2 \kappa^2$, 
\begin{equation}
\chi(w)=2\cos\left(\frac{\pi+\arccos\left(\tanh\left(\frac{a b^2 w}{\kappa^2}\right)\right)}{3}\right).
\end{equation}
that connects the minima $m\pi/2$ to $(m+1)\pi/2$, with $m$ integer. The  warp function equation given by Eq. \eqref{fo2Ex2} is $A^\prime = -a\chi/3$, therefore
\be\label{A20}
A(w)=-\frac{2a}{3} \bigintsss^w 
\cos\left(\frac{\pi+\arccos\left(\tanh\left(\frac{a b^2 \tilde w}{\kappa^2}\right)\right)}{3}\right)
d\tilde w.  
\ee
In Fig. \ref{figWarp21}, we depict the warp factor $e^{2A(w)}$ for $a=1$, and $\kappa=1$, and $b=1/2, 1, 2$. It shows that the thickness of the configuration decreases as $b$ increases.

\begin{figure}[ht]
    \centering
\includegraphics[width=0.40\textwidth]{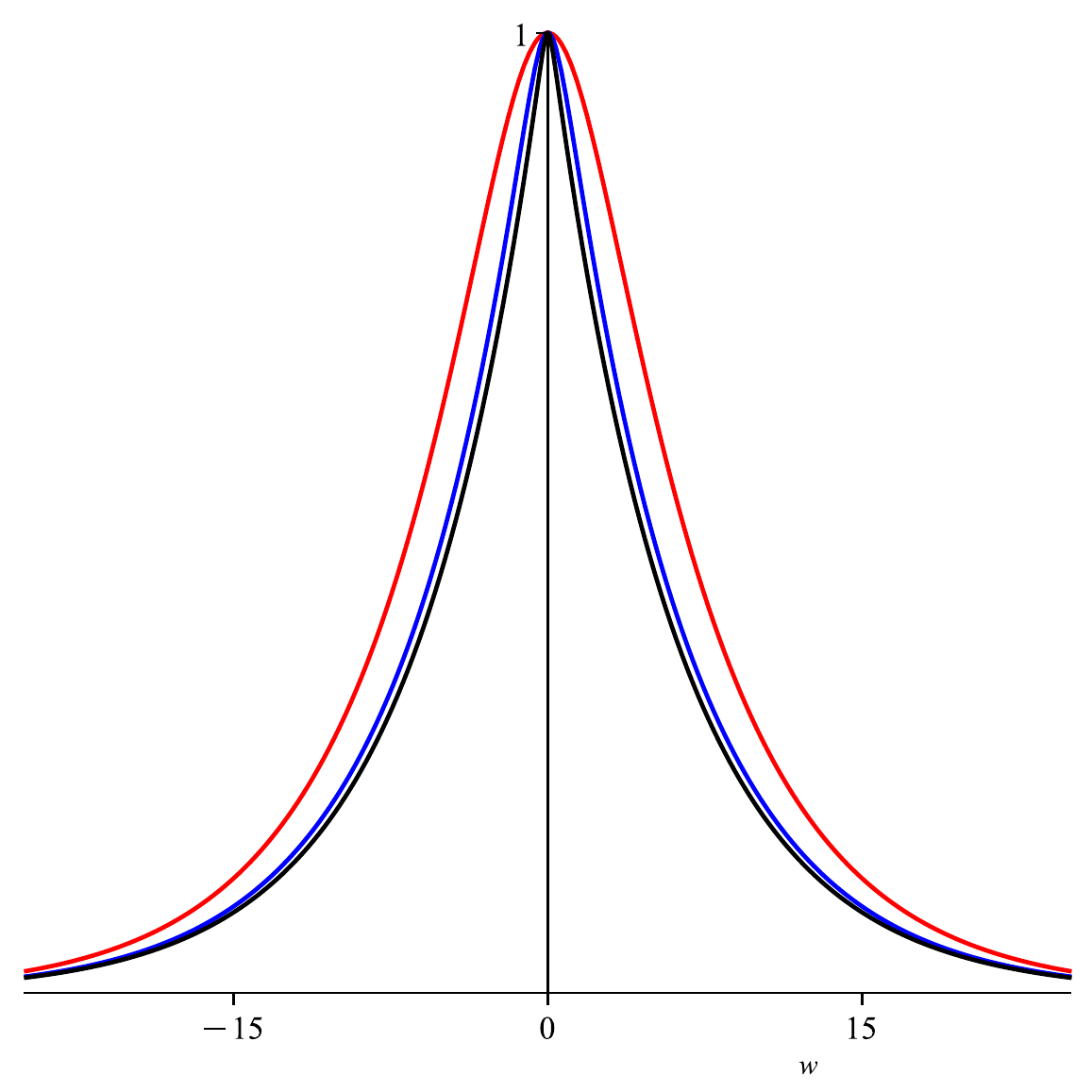}
\includegraphics[width=0.40\textwidth]{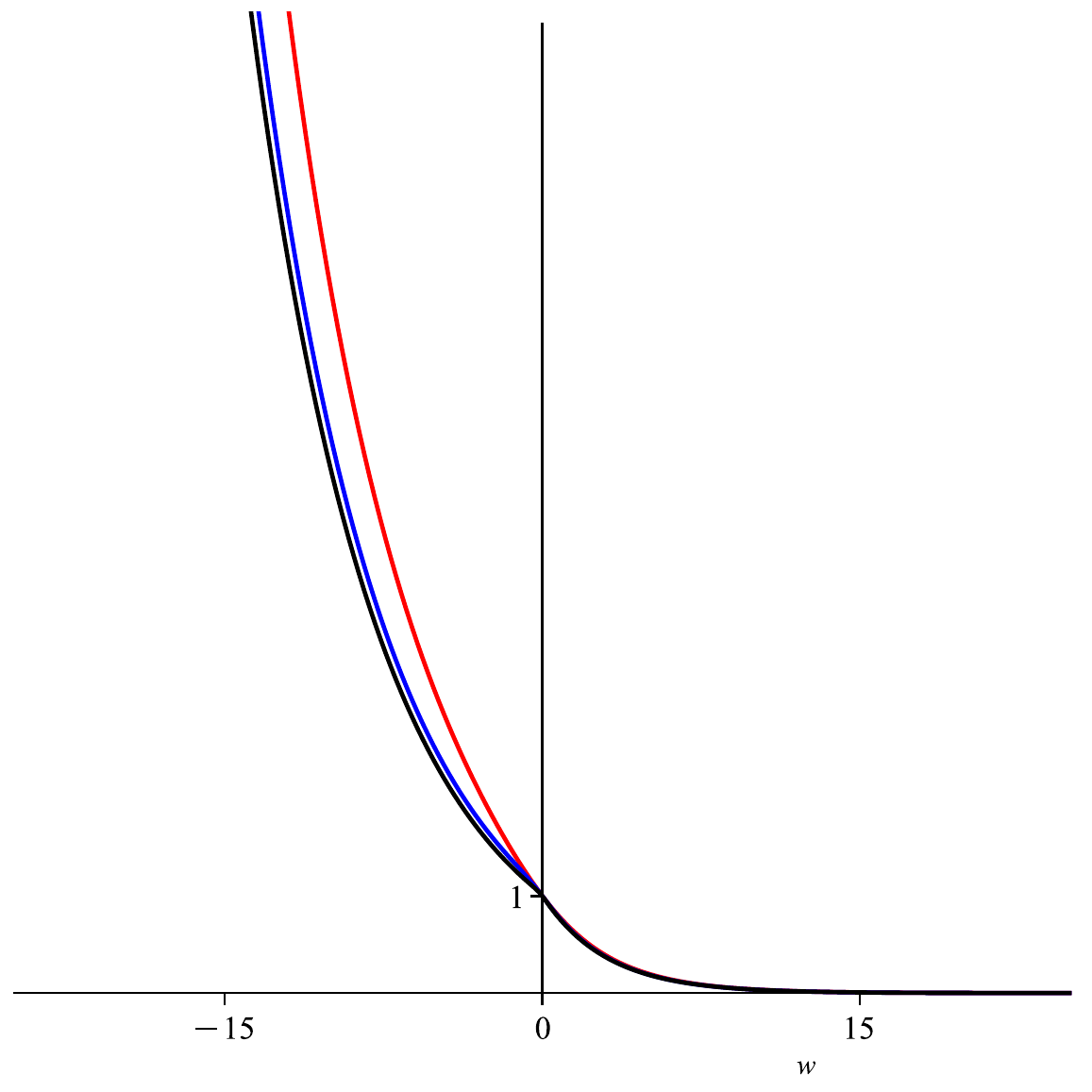}
     \caption{The warp factor using $A_{I}$ (left) and $A_{II}$ (right) given by Eqs. \eqref{warpBi} and \eqref{warpBii} for $\alpha=4\alpha_1.$ 
     We take $a=1$, $\kappa=1$ and $b=1/2$ (red), $1$ (blue), and $2$ (black).}
    \label{figWarp2}
\end{figure}

The second possibility is for $\alpha=4\alpha_1= 9/(a^2 \kappa^2)$. In this case, we have two kinds of solutions 
\begin{eqnarray}
\chi_I(w)&=&\cos\left(\frac{\pi+\arccos\left(\tanh\left(\frac{2a b^2 w}{\kappa^2}\right)\right)}{3}\right), \\
\chi_{II}(w)&=&\cos\left(\frac{\arccos\left(\tanh\left(\frac{2a b^2 w}{\kappa^2}\right)\right)}{3}\right).
\end{eqnarray}
For $b^2>\kappa^2/2$, these solutions connect two minima. The corresponding warp functions are
\begin{eqnarray}\label{warpBi}
A_I(w)&=&-\frac{a}{3} \bigintsss^w 
\cos\left(\frac{\pi+\arccos\left(\tanh\left(\frac{2a b^2 \tilde w}{\kappa^2}\right)\right)}{3}\right)
d\tilde w, \\ 
\label{warpBii}
A_{II}(w)&=&-\frac{a}{3} \bigintsss^w 
\cos\left(\frac{\arccos\left(\tanh\left(\frac{2a b^2 \tilde w}{\kappa^2}\right)\right)}{3}\right)
d\tilde w.
\end{eqnarray}
In Fig. \eqref{figWarp2},  we display  the warp factors $e^{2A_I(w)}$ and $e^{2A_{II}(w)}$ for $a=1$, and $\kappa=1$, and $b =1/2, 1, 2$. Notice that only solutions with $A_{I}$ arise connecting $AdS$ geometries, with the thickness decreasing as $b$ increases. 

\section{Summary and conclusion}
\label{summary}

In this paper, we have tackled $2$-brane solutions in four-dimensional $EGB$ gravity. First, by solving the $EGB$ field equations, 
which are the same in both formulations of $4D$ $EGB$, for a flat $2$-brane ansatz in the absence of matter fields, we found that its worldvolume is described by two copies of $AdS_{4}$, with the cosmological constant entirely sourced by a positive GB coupling constant, $\alpha>0$. As expected, this solution has no analog within GR, once the limit $\alpha\rightarrow 0$ is not well defined.
 
In order to capture some properties of the aforementioned solution, we added a massive probe scalar field and then solved the corresponding Klein-Gordon equation in the bulk metric. However, since the $AdS_{4}$ metric possesses a singularity at the horizon in the usual Poincarè coordinates, we have obtained its description in Gaussian null coordinates which, in turn, has made it clear that $AdS_{4}$ can be viewed as a warped product between $AdS_{2}$ and $\mathcal{H}^{2}$, and also its regularity at the horizon. By solving the wave equation in this non-singular coordinate system, we were able to find the normalizable and non-normalizable modes that play an important role in the choice of the boundary conditions to eliminate the contributions stemming from the boundary. In fact, we have found that only the normalizable modes lead to the vanishing of the boundary term. Such a requirement is attained by imposing a quantization condition on the separation constant that appears in Eqs. \eqref{qc} and \eqref{qc1}.

Furthermore, we have investigated the system in the presence of a source scalar field having self-interaction, searching for the possibility to construct braneworld solutions within the new $AdS_4$ geometry under the modified GB gravity. In this case, after including an auxiliary function $W(\Phi)$, we have been able to introduce a first-order procedure, with both the warp function $A$ and the scalar field being controlled by the first-order equations \eqref{fo1} and \eqref{fo2}, which solve the equations of motion when the potential has the form displayed in \eqref{potW}. These general results show that the GB modification nicely contributes to the change of the warp factor, making the configuration represent a thick brane. We have illustrated the results with several distinct models, in particular, with the {\it type-I} model which supports analytical solutions for both the scalar field and the warp factor, so it can serve as a good model to be considered for extensions that add other modifications. The presence of $W(\Phi)$ and the accompanying first-order equations suggest that we investigate issues related to fake supergravity \cite{Nunes} and the first-order Hamilton–Jacobi equations \cite{Town,Riet}, to see if novelties appear in the present scenario. 

The models studied in the present work can be enlarged to include other fields, in particular, another scalar field to see if the brane can support internal modification, in a way similar to the cases considered in Refs. \cite{BB,BB1}, where mechanisms to control the internal structure of the brane were developed. We can also study the entrapment of fermions and other fields in the brane; see, e.g., Refs. \cite{Liu,Adalto,Hott}. It is also possible to introduce other geometric modifications, similar to the ones considered in Refs. \cite{Lobao,Lobao1} in the five-dimensional spacetime. 
Another interesting possibility is to change the metric in Eq. \eqref{metric} to the more general case, having the bent brane profile
\bea
ds^2=e^{2A(w)}g_{ab}dx^adx^b-dw^2,
\eea
where $g_{ab}$ may now engender nontrivial three-dimensional geometry. This may make a direct connection to the study described in Refs. \cite{Olmo}, in which one considers Born-Infeld extension of general relativity formulated in metric-affine spaces in $2+1$ dimensions to find analytical solutions and also, to the old results described in Ref. \cite{2brane1,2brane2}, in which the authors discovered that in the case of the bulk $AdS_4$, a class of brane-localized black hole metrics is possible. 
These and other specific issues are currently under consideration, and we hope to report on them in the near future.

\appendix
\section{Regularized Kaluza-Klein reduction for $4D$ $EGB$}
\label{app}
\hspace{0.5cm}
This Appendix is devoted to deriving the field equations of regularized $4D$ $EGB$ via Kaluza-Klein dimensional reduction. To begin with, let us consider the ansatz (\ref{metric}), which describes a brane-like metric. Hence upon Kaluza-Klein dimensional reduction, (\ref{GBaction}) reduces to the effective action below
\begin{eqnarray}
  \nonumber  S_{_{\mbox{eff}}}&=&\int\,d^{\mathcal{D}-1}x\int\, dw \,e^{(\mathcal{D}-1)A}\bigg[\frac{(\mathcal{D}-1)(\mathcal{D}(A^{\prime})^2 +2A^{\prime\prime})}{2\kappa^2}-\alpha(\mathcal{D}-1)(\mathcal{D}-2)(\mathcal{D}-3)\times\\
&\times&\bigg(\mathcal{D}(A^{\prime})^4
    +4A^{\prime\prime}(A^{\prime})^2\bigg)\bigg].
    \label{eff}
\end{eqnarray}
For the sake of convenience, let us redefine $\mathcal{D}=4+\epsilon$, then the former equation becomes
\begin{eqnarray}
  \nonumber  S_{_{\mbox{eff}}}&=&\int\,d^{3+\epsilon}x\int\, dw \,e^{(3+\epsilon)A}\bigg[\frac{(3+\epsilon)\left((4+\epsilon)(A^{\prime})^2 +2A^{\prime\prime}\right)}{2\kappa^2}-\alpha(3+\epsilon)(2+\epsilon)(1+\epsilon)\times\\
  &\times&\bigg((4+\epsilon)(A^{\prime})^4
    +4A^{\prime\prime}(A^{\prime})^2\bigg)\bigg].
\end{eqnarray}
In order to have a non-trivial contribution to the \textit{GB} term by taking the limit $\mathcal{D}\rightarrow 4$ or $\epsilon\rightarrow 0$, we must add a counterterm to the former action. Its explicit form is given by
\begin{equation}
    S_{_{\mbox{c.t.}}}=\alpha \int\,d^{3}x\int\,dw\, e^{3A}\mathcal{G}=24\alpha \int\,d^{3}x\int\,dw\, e^{3A}\left((A^{\prime})^4
    +A^{\prime\prime}(A^{\prime})^2\right).
    \label{coun}
\end{equation}
Therefore, by adding (\ref{coun}) to the action (\ref{eff}), the regularized action, after proceeding with the redefinitions $\alpha\rightarrow \dfrac{\alpha}{\mathcal{D}-4}=\dfrac{\alpha}{\epsilon}$ and $\mathcal{D}\rightarrow 4$ or $\epsilon\rightarrow 0$, looks like
\begin{eqnarray}
     S_{_{\mbox{reg}}}=\int\,d^3 x \int\,dw\,\frac{3}{2\kappa^2}e^{3A}\left[4(A^{\prime})^2 +2A^{\prime\prime}\right]+2\alpha\int\,d^3 x \int\,dw\,e^{3A}(A^{\prime})^{4},
     \label{reg}
\end{eqnarray}
where we got rid of the boundary terms to find the aforementioned regularized action in four dimensions. Note that the three-dimensional section of the bulk can be omitted since the effective dynamics of the warped function $A=A(w)$ is restricted to the ``extra'' dimension coordinate only; thereby, the only effective dynamical degree of freedom is $A$. 

Varying (\ref{reg}) with respect to the scalar field $A$ and disregarding the boundary terms we are able to find 
\begin{eqnarray}
    \delta  S_{_{\mbox{reg}}}= 6\int d^3 x \int dw\,e^{3A}\left[\frac{1}{2\kappa^2}\left(3(A^{\prime})^2 +2A^{\prime\prime}\right)-\alpha\left(3(A^{\prime})^4 +4A^{\prime\prime}(A^{\prime})^2\right)\right]\delta A.\;\;
    \label{delta1}
\end{eqnarray}
Now, varying the matter action with respect to $A$, considering the scalar field of the subsection  (\ref{subsection}) as the matter source, we obtain
\begin{equation}
    \delta S_{m}=3\int d^3 x \int dw\,e^{3A}\left[\frac{1}{2}\Phi^{\prime 2}+V(\Phi)\right]\delta A.
    \label{delta2}
\end{equation}
Defining the total action as $S=S_{_{\mbox{reg}}} +S_{m}$ and using Eqs. \eqref{delta1} and \eqref{delta2}, the variational principle tells us that
\begin{eqnarray}
    0&=&\delta S\\
   \nonumber &=&\frac{3}{\kappa^2}\int d^3 x \int dw\,e^{3A}\bigg[\left(3(A^{\prime})^2 +2A^{\prime\prime}\right)-2\alpha\kappa^2\left(3(A^{\prime})^4 +4A^{\prime\prime}(A^{\prime})^2\right)+\\
&+&    \kappa^2 \left(\frac{1}{2}\Phi^{\prime 2}+V(\Phi)\right)\bigg]\delta A,
\end{eqnarray}
providing the following field equation
\begin{equation}
    \kappa^2 \left(\frac{1}{2}\Phi^{\prime 2}+V(\Phi)\right)=-3(A^{\prime})^2 -2 A^{\prime\prime}+2\alpha\kappa^2 \left(3(A^{\prime})^4 + 4A^{\prime\prime}(A^{\prime})^2 \right),
    \label{A9}
\end{equation}
which coincides with Eq. \eqref{oneE} found via Glavan and Lin procedure \cite{newGB}. Now, to find the other field equation, we must trace Eq. \eqref{fe}\footnote{As discussed in Section \ref{general}, the trace of Eq. \eqref{fe} is well-defined in the limit $\mathcal{D}\rightarrow 4$ or $\epsilon\rightarrow 0$ and for the Kaluza-Klein regularization method such an equation holds, see \cite{Lu}. } and take the limit $\mathcal{D}\rightarrow 4$ or $\epsilon\rightarrow 0$, then we arrive at
\begin{equation}
    \kappa^2\left(\Phi^{\prime 2}+4V(\Phi)\right)=-6\left[2(A^{\prime})^2 +A^{\prime\prime}\right]+24\alpha\kappa^2\left((A^{\prime})^4+(A^{\prime})^2 A^{\prime\prime}\right),
\end{equation}
combining with Eq. \eqref{A9}, one has
\begin{equation}
    -3(A^{\prime})^2+6\alpha\kappa^2(A^{\prime})^4= \kappa^2\left(-\frac{1}{2}\Phi^{\prime 2}+V(\Phi)\right),
\end{equation}
which is the same equation (\ref{twoE}) found via Glavan and Lin procedure \cite{newGB}. Therefore, for the ansatz \eqref{metric}, the $4D$ $EGB$ performing the Kaluza-Klein regularization shares the same field equations as the original formulation \cite{newGB}.
 
\section*{Acknowledgments}
\hspace{0.5cm}

This work is partially supported by Conselho Nacional de Desenvolvimento Científico e Tecnológico, Grants  303469/2019-6 (DB), 310994/2021-7 (RM), 303777/2023-0 (AYP), and 307628/2022-1 (PJP) and by Paraiba State Research Foundation, Grants 0015/2019 (DB and AYP), 0003/2019 (RM) and 150891/2023-7 (AYP and PJP).

\end{document}